%% file: main.tex
\documentclass[journal]{IEEEtran}
\usepackage{multirow}
\usepackage{mathtools}
\usepackage{amsmath,amssymb,amsfonts}
\usepackage{graphicx}
\usepackage{textcomp}
\usepackage{xcolor}
\usepackage{subfigure}
\usepackage[letterpaper, left=0.65in, right=0.65in, bottom=1in, top=0.70in]{geometry}

\def\BibTeX{{\rm B\kern-.05em{\sc i\kern-.025em b}\kern-.08em
    T\kern-.1667em\lower.7ex\hbox{E}\kern-.125emX}}
\usepackage{xcolor}
\usepackage[linesnumbered,ruled,vlined]{algorithm2e}
\usepackage{listings}

\SetCommentSty{mycommfont}
\SetKwInput{KwInput}{Input}   
\SetKwInput{KwOutput}{Output} 

\definecolor{GreenForest}{rgb}{0.09, 0.45, 0.27}
\usepackage{soul}
\usepackage{balance}
\usepackage{cite}
\usepackage[nolist]{acronym}
\usepackage{acronym}

\usepackage{comment}
\usepackage{amsmath}
\usepackage{steinmetz}

\makeatletter
\newcommand{\ie}{\emph{i.e.}\@ifnextchar.{\!\@gobble}{}}
\newcommand{\eg}{\emph{e.g.}\@ifnextchar.{\!\@gobble}{}}
\newcommand{\etc}{etc\@ifnextchar.{}{.\@}}
\makeatother

\begin{document}
\input{acronyms}

\bstctlcite{IEEEexample:BSTcontrol}
    \title{Optimized Non-Uniform Pilot Pattern for \ac{OFDM} Sensing}
  \author{Amir Bouziane, and H\"{u}seyin Arslan,~\IEEEmembership{Fellow,~IEEE}

\thanks{The authors are with the Department of Electrical and Electronics Engineering, Istanbul Medipol University, Istanbul, 34810, Turkey (e-mail: bouziane.amir@std.medipol.edu.tr;  huseyinarslan@medipol.edu.tr).}
}

\IEEEpeerreviewmaketitle
\maketitle

\footnote{\textit{Notation:} Bold uppercase $\mathbf{A}$ and lowercase $\mathbf{a}$ denote matrices and column vectors, respectively. $(\cdot)^T$ and $(\cdot)^H$ denote the transpose and conjugate transpose (Hermitian) operators. $\mathrm{diag}(\cdot)$ constructs a diagonal matrix from a vector or list of scalars. $|\cdot|$ denotes the magnitude of a scalar or the cardinality of a set. $\mathbb{I}(\cdot)$ denotes the indicator function. $\mathbb{C}^{M \times N}$ represents the space of $M \times N$ complex matrices. $\mathcal{CN}(\mathbf{0}, \sigma^2\mathbf{I})$ denotes the circularly symmetric complex normal distribution with zero mean and covariance $\sigma^2\mathbf{I}$.}

\begin{abstract}
Standard periodic pilot patterns in orthogonal frequency division multiplexing (OFDM) systems induce severe delay-domain grating lobes, compromising radar sensing. This paper proposes a two-stage framework to design non-periodic pilot patterns that minimize the peak sidelobe level (PSL) \textcolor{black}{while strictly enforcing communication anchor constraints.} We \textcolor{black}{solve this combinatorial problem using a low-complexity} hybrid greedy-stochastic cyclic coordinate descent (SCCD) algorithm. This approach \textcolor{black}{shatters cyclic periodicities to suppress deterministic grating lobes beneath the impassable data-to-pilot interference (DPI) noise floor. System-level evaluations demonstrate the performance of the proposed design in resolving the sensing-communication trade-off, showing improved range root mean square error (RMSE) without degrading the primary communication bit error rate (BER).}
\end{abstract}



\section{Introduction}
\label{sec:intro}

\ac{ISAC} has emerged as a central element of the \ac{6G} vision, enabling wireless systems to sense and communicate using the same radio resources \cite{gonzalez2024integrated, liu2024senscap, lee20226g}. Among the available waveform candidates, \ac{OFDM} remains the most practical foundation for \ac{ISAC} due to its established use in \ac{5G}~NR and Wi-Fi, its flexible subcarrier structure, and its well-characterized sensing behavior under hardware impairments \cite{keskin2023monostatic, liu2025cp, shi2025ofdm}.

Despite these advantages, \ac{OFDM} was not originally designed with high-resolution sensing in mind. A key limitation stems from its pilot structure. Standard communication protocols employ periodic pilot placement for straightforward channel estimation \cite{coleri2002channel}, but in sensing applications these periodically spaced tones form a frequency-domain comb whose inverse transform yields a delay-domain pattern dominated by regularly spaced grating lobes \cite{lu2024sensing}. These lobes appear as strong ghost targets and can mask weaker reflections, creating a fundamental conflict between communication-oriented pilot design and reliable ranging performance.

Several recent works attempt to improve \ac{OFDM}-based sensing by modifying the waveform, subcarrier mapping, or symbol constellation \cite{geiger2025joint, du2024reshaping, shi2025ofdm}. While such approaches can suppress sidelobes, they often require changes to the modulation format or receiver processing, limiting compatibility with standardized \ac{OFDM} systems. A more practical alternative \textcolor{black}{is to retain the \ac{OFDM} structure and optimize only the pilot locations. The design of spectrally-constrained sequences with low sidelobes is a classical problem, yielding powerful algebraic difference sets and convex optimization frameworks \cite{R1,R2,R3}. However, these theoretical bounds generally assume unconstrained grid dimensions and offline synthesis. In standardized \ac{3GPP} \ac{OFDM} numerologies, the strict existence conditions for perfect algebraic sequences are not met, rendering the placement task a combinatorial NP-hard problem. Furthermore, seamless integration into existing \ac{3GPP} frameworks necessitates and prefer practical, hardware-friendly solutions that can be readily adopted by the standard}

This work develops a pilot-design approach that directly targets the structural origin of delay sidelobes. We analyze the role of cyclic pairwise differences between active tones and show that the distribution of these differences rather than the individual pilot indices determines the shape of the periodic ambiguity function. Guided by this observation, we design a hybrid optimization method that first forms a constructive low-sidelobe pattern and then refines it using stochastic cyclic coordinate descent to escape the local minima that limit purely greedy methods.

The main contributions of this paper are as follows.
\textcolor{black}{First, we link cyclic difference multiplicities to the ambiguity function to mathematically explain periodic grating lobes. Second, we update the ISAC model to include the data-to-pilot interference (DPI) noise floor. Third, we propose a low-complexity Hybrid Greedy-SCCD algorithm that optimizes discrete pilot placement while strictly enforcing communication anchors. Finally, we demonstrate the performance of the proposed method by showing it range \ac{RMSE} and communication \ac{BER} under practical assumptions.}


\section{System Model}
\label{sec:system_model}

We consider an \ac{OFDM} probing waveform with $N$ equispaced subcarriers. Let
\begin{equation}
    \mathbf{s}_p \in \{0,1\}^N
\end{equation}
denote the pilot-activation vector, where the support
\begin{equation}
    \mathcal{P} = \{ k : [\mathbf{s}_p]_k = 1 \}, \qquad |\mathcal{P}| = K,
\end{equation}
specifies the active pilot tones. Each active tone carries a unit-magnitude pilot so that the effect of tone selection is isolated from power allocation and modulation.

The baseband time-domain probing sequence is obtained through the standard \ac{OFDM} modulation operation \cite{weinstein2003data,nee2000ofdm}:
\begin{equation}
    \mathbf{x} = \mathbf{F}^{H}\mathbf{s}_p,
\end{equation}
where $\mathbf{F}$ is the unitary $N$-point \ac{DFT} matrix with elements
\begin{equation}
    [\mathbf{F}]_{m,n}=\frac{1}{\sqrt{N}} e^{-j2\pi mn/N}.
\end{equation}

A \ac{CP} of length $N_{\mathrm{cp}}$ is appended in practice to ensure that the time-domain channel convolution becomes circular, enabling subcarrier-wise diagonalization after \ac{DFT} processing \cite{li2006orthogonal}. Throughout the analysis, we assume perfect \ac{CP} removal and focus on the $N$-sample useful block.

\subsection{Sensing Channel and Received Signal}

We adopt a monostatic \ac{OFDM} sensing model, commonly used in radar signal processing and \ac{OFDM} based radar studies \cite{braun2014ofdm, skolnik2008radar}. Let the scene contain $L$ point targets characterized by complex gains $\alpha_l$ and integer delays $\tau_l$ in samples. Under an \ac{OFDM} symbol, the received baseband signal is modeled as
\begin{equation}
    \mathbf{y}
    = \sum_{l=0}^{L-1} \alpha_l \mathbf{J}_{\tau_l}\mathbf{x} + \mathbf{z},
\end{equation}
where $\mathbf{z}\sim\mathcal{CN}(0,\sigma^2\mathbf{I})$ represents additive noise. \textcolor{black}{The Signal-to-Noise Ratio (SNR) is defined as $\mathrm{SNR} = E_s/N_0$, where $E_s$ represents the average transmitted energy per symbol and $N_0 = \sigma^2$ denotes the one-sided noise power spectral density.} The matrix $\mathbf{J}_{\tau}$ is the circular delay operator defined as
\begin{equation}
    (\mathbf{J}_{\tau})_{m,n}
    = \delta[(m-n+\tau)\bmod N].
\end{equation}

Circular shift matrices form a circulant family and are diagonalized by the \ac{DFT} \cite{gray2006toeplitz}. This property is central to the structure of the ambiguity function derived below.
\subsection{Matched-Filter Output and Ambiguity Function}

Range processing is performed by correlating the received signal with circularly shifted versions of the transmitted sequence:
\begin{equation}
    r(\tau)=\mathbf{x}^{H}\mathbf{J}_{\tau}\mathbf{y}
           = \sum_{l=0}^{L-1} \alpha_l \textcolor{black}{\Psi_{\mathrm{total}}}(\tau+\tau_l)
             + \tilde{z}(\tau),
\end{equation}
where $\tilde{z}(\tau)=\mathbf{x}^{H}\mathbf{J}_\tau\mathbf{z}$ is filtered noise.
\
The \textcolor{black}{total} delay-only periodic ambiguity function (AF) of the probing sequence \cite{skolnik2008radar} is \textcolor{black}{$\Psi_{\mathrm{total}}(\tau) = \mathbf{x}^{H}\mathbf{J}_{\tau}\mathbf{x}$}. 
\textcolor{black}{Because $\mathbf{x} = \mathbf{F}^{H}(\mathbf{s}_p + \mathbf{d})$, this expands into deterministic and stochastic components: 
\begin{equation}
    \Psi_{\mathrm{total}}(\tau) = \Psi_{pp}(\tau) + \Psi_{dd}(\tau) + \Psi_{pd}(\tau) + \Psi_{dp}(\tau).
\end{equation} Since the communication data $\mathbf{d}$ comprises independent random variables, its autocorrelation $\Psi_{dd}(\tau)$ creates an impassable stochastic noise floor, theoretically bounded at approximately $10 \log_{10}(1/N)$. Consequently, our design objective is not to achieve the absolute zero-sidelobe bounds of isolated sequences, but to strictly suppress the deterministic pilot grating lobes ($\Psi_{pp}$) beneath this unavoidable communication data floor.}

Using \textcolor{black}{$\mathbf{x}_p=\mathbf{F}^{H}\mathbf{s}_p$ for the deterministic pilot component} and the diagonalization property of circulant matrices \cite{gray2006toeplitz}, \textcolor{black}{$\mathbf{F}\mathbf{J}_{\tau}\mathbf{F}^{H} = \mathbf{D}_{\tau}$}:
\begin{equation}
    \mathbf{D}_{\tau}
    = \mathrm{diag}\!\big(1, e^{j2\pi\tau/N}, \dots, e^{j2\pi(N-1)\tau/N}\big),
\end{equation}
we obtain \textcolor{black}{the deterministic pilot AF}:
\begin{equation}
    \textcolor{black}{\Psi_{pp}}(\tau)
    = \mathbf{s}_p^{H} \mathbf{D}_{\tau}\mathbf{s}_p
    = \sum_{k\in\mathcal{P}} e^{j2\pi k \tau/N}.
\end{equation}
Thus, the \textcolor{black}{deterministic} ambiguity structure \textcolor{black}{dictating the grating lobes} depends solely on the selected pilot set $\mathcal{P}$.

\textcolor{black}{For a coherent processing interval of $M$ \ac{OFDM} symbols, the integrated ambiguity function is the sum of the individual symbol responses: 
\begin{equation}
    R_{\mathrm{total}}(\tau) = \sum_{m=1}^{M} \Psi_{\mathrm{total}}^{(m)}(\tau).
\end{equation}
Because the pilot pattern is static across the interval, the deterministic pilot component integrates coherently, with its power scaling quadratically ($M^2$). Conversely, the independent data payload integrates incoherently, with its power scaling linearly ($M$).}
\textcolor{black}{\subsection{Communication Channel Model}
To evaluate the communication performance, we define a frequency-selective multipath channel. The discrete-time impulse response is modeled as \cite{4607239}:
\begin{equation}
    h[n] = \sum_{p=1}^{P} \beta_p \delta[n-\nu_p],
\end{equation}
where $P$ is the number of resolvable paths, and $\beta_p$ and $\nu_p$ are the complex gain and integer delay of the $p$-th path, respectively. The baseband received signal at the communication user, after cyclic prefix removal, is expressed in vector form as :
\begin{equation}
    \mathbf{y}_c = \sum_{p=1}^{P} \beta_p \mathbf{J}_{\nu_p}\mathbf{x} + \mathbf{z}_c,
\end{equation}
where $\mathbf{J}_{\nu_p}$ is the circular delay operator defined in (6), and $\mathbf{z}_c \sim \mathcal{CN}(\mathbf{0}, \sigma_c^2\mathbf{I})$ is the receiver noise. The frequency-domain channel is estimated using the active pilot set $\mathcal{P}$ to equalize the data payload $\mathbf{d}$.}

\subsection{Peak Sidelobe Level Metric}

Delay sidelobes are quantified using the \ac{PSL}, a standard radar performance metric defined \cite{skolnik2008radar}. Let
\begin{equation}
    d(\tau)=\min\{\tau,\, N-\tau\},
\end{equation}
and define the sidelobe region
\begin{equation}
    \mathcal{T}_{\mathrm{sl}}
    = \{ \tau : \tau_{\min} \le d(\tau)\le \tau_{\max}\}.
\end{equation}

The \ac{PSL} is then
\begin{equation}
    \mathrm{PSL} =
    \max_{\tau\in\mathcal{T}_{\mathrm{sl}}}
    \frac{|\Psi(\tau)|^2}{|\Psi(0)|^2}.
\end{equation}
Since $\Psi(0)=K$, the normalization satisfies $|\Psi(0)|^2=K^2$.

\section{Difference-Set Representation of the Ambiguity Function}
\label{sec:diffset}

\subsection{Difference Multiplicity Function}

The delay-domain \ac{AF} is determined by the pairwise differences between active tones. For $\mathcal{P}\subset\{0,\dots,N-1\}$, define the cyclic difference multiplicity
\begin{equation}
    \lambda(d)
    =
    \sum_{n\in\mathcal{P}}
    \sum_{m\in\mathcal{P}}
    \mathbb{I}\!\left(m-n \equiv d \; (\mathrm{mod}\; N)\right),
     d \in \{0,\dots,N-1\},
\end{equation}
where $\mathbb{I}(\cdot)$ \textcolor{black}{denotes the indicator function, which evaluates to 1 if the cyclic congruence condition is satisfied, and 0 otherwise.} By construction,
\begin{equation}
    \sum_{d=0}^{N-1} \lambda(d) = K^2,
\end{equation}
and $\lambda(0) = K$ counts the trivial zero differences.
\begin{figure}[t]
\centering
\includegraphics[width=0.5\textwidth]{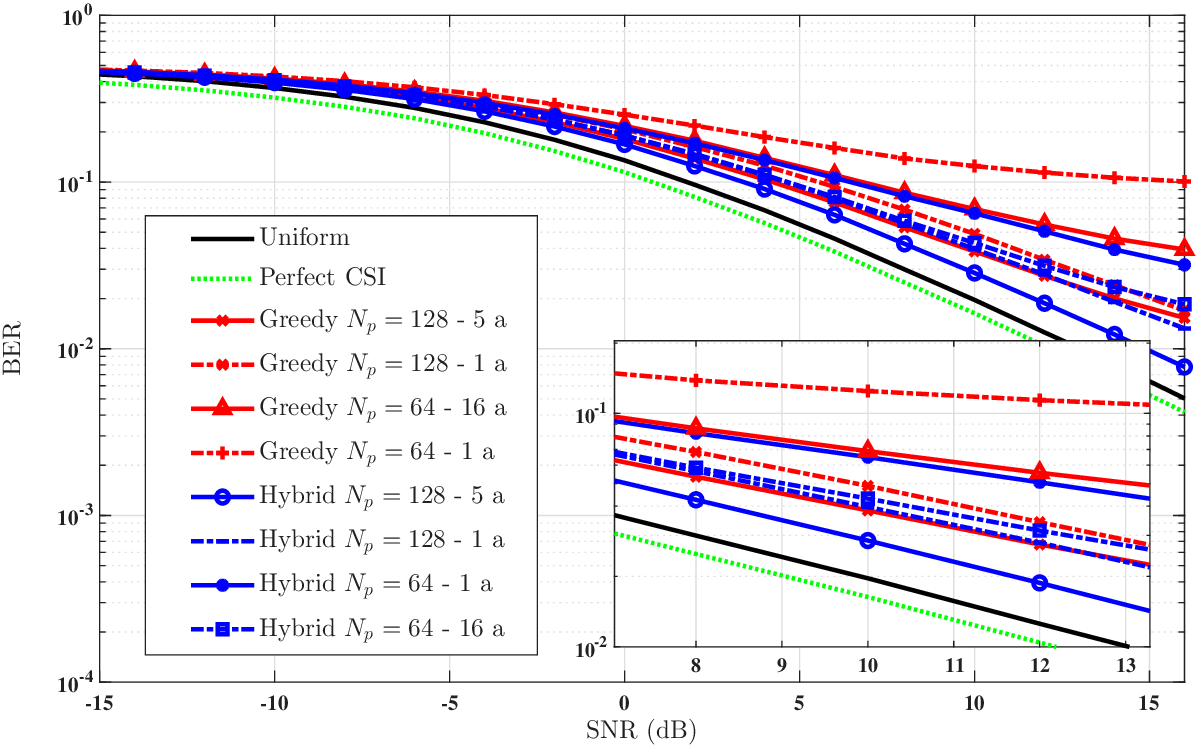}
\caption{BER versus SNR over a frequency-selective Rayleigh fading channel.}
\label{fig:ber}
\end{figure}


\subsection{Ambiguity Function in Terms of Differences}

Expanding the squared magnitude of the \ac{AF} yields
\begin{align}
    |\Psi(\tau)|^2
    &= \left| \sum_{m\in\mathcal{P}} e^{j\frac{2\pi}{N}m\tau} \right|^2 \\
    &= \sum_{m\in\mathcal{P}} \sum_{n\in\mathcal{P}}
       e^{j\frac{2\pi}{N}(m-n)\tau} \\
    &= \sum_{d=0}^{N-1}
       \lambda(d) e^{j\frac{2\pi}{N} d \tau}.
\end{align}
Equivalently,
\begin{equation}
    |\Psi(\tau)|^2
    = K + \sum_{d\neq 0} \lambda(d) e^{j\frac{2\pi}{N} d \tau}.
\end{equation}
Thus, $\{|\Psi(\tau)|^2\}_{\tau=0}^{N-1}$ is the $N$-point \ac{DFT} of the nonnegative sequence $\{\lambda(d)\}_{d=0}^{N-1}$. For fixed $(N,K)$, the entire delay-domain \ac{AF} is therefore encoded in the cyclic difference multiplicities $\lambda(d)$. Intuitively, highly non-uniform difference multiplicities lead to stronger constructive phasor alignment for certain lags $\tau$, and hence to high sidelobes. Patterns whose differences are more evenly spread in $d$ tend to exhibit flatter sidelobes.

\subsection{Stage 1: Constructive Sidelobe Minimization (CSM)}
\textcolor{black}{While an unconstrained greedy search minimizes the radar PSL, it mathematically clusters pilots in contiguous subcarriers, which prevents the communication receiver from accurately interpolating the frequency-selective channel. To explicitly manage this sensing-communication trade-off, we enforce a structural constraint before the optimization begins. A minimum subset of $N_{anc}$ pilot anchors are  scattered across the grid to guarantee strict wideband channel estimation coverage. This ensures that the channel sampling frequency satisfies the Nyquist criterion relative to the channel coherence bandwidth, preventing the communication outages inherent to unconstrained sensing-only optimization}
\begin{figure}[t]
\centering
\includegraphics[width=0.5\textwidth]{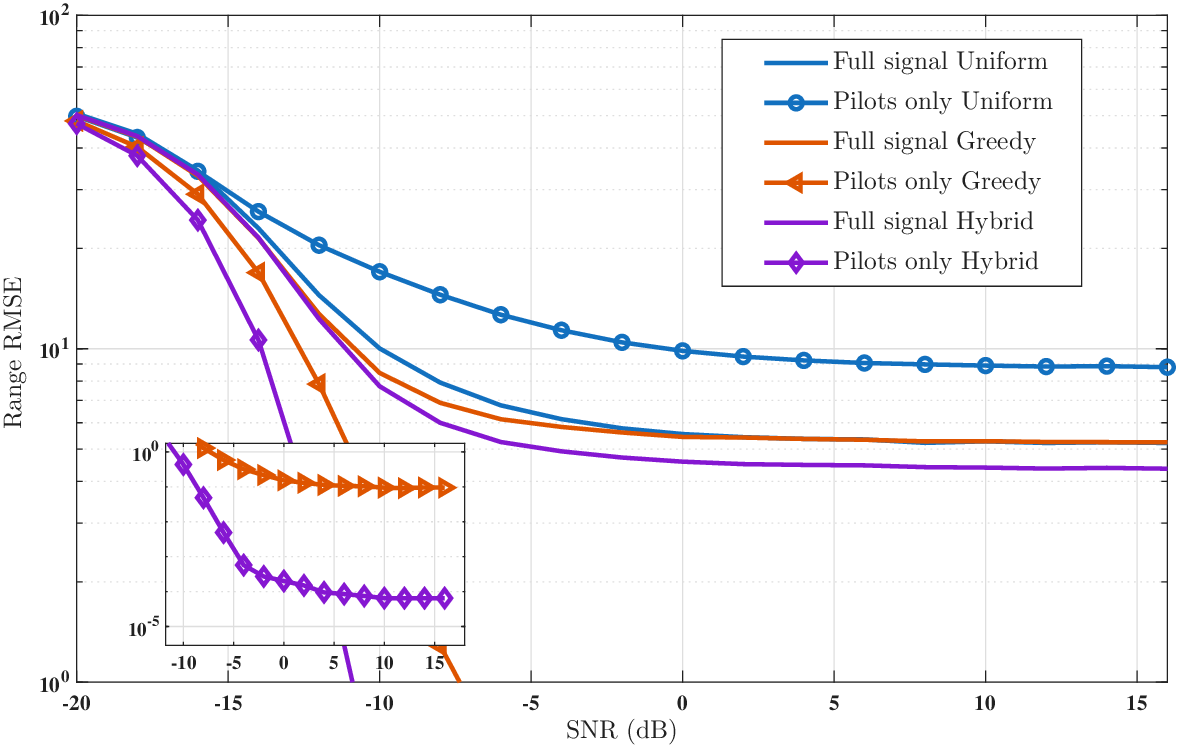}
\caption{Range RMSE versus SNR.}
\label{fig:rmse}
\end{figure}
\textcolor{black}{Let $\mathcal{P}_{anc}$ be this initial anchor set of size $N_{anc}$. The remaining $K - N_{anc}$ free pilots are then iteratively selected to minimize the instantaneous \ac{PSL}.} Let $\mathcal{P}_{k-1}$ be the set of $k-1$ selected pilots \textcolor{black}{(starting with $\mathcal{P}_{N_{anc}} = \mathcal{P}_{anc}$)}. The $k$-th pilot index $n^\star$ is chosen from the available set $\mathcal{U} = \{0, \dots, N-1\} \setminus \mathcal{P}_{k-1}$ such that:
\begin{equation}
    n^\star = \underset{n \in \mathcal{U}}{\mathrm{argmin}} \left( \max_{\tau \in \mathcal{T}_{\mathrm{sl}}} \left| \Psi\textcolor{black}{_{pp}}(\tau; \mathcal{P}_{k-1} \cup \{n\}) \right| \right).
\end{equation}
This step is repeated until $|\mathcal{P}| = K$. While computationally efficient ($O(K \cdot N)$), this constructive approach is prone to stagnation in local minima, \textcolor{black}{especially under strict anchor constraints}, as decisions made in early iterations are never revisited.

\begin{algorithm}[t]
\caption{Hybrid Pilot Pattern Design}
\label{alg:hybrid_pilot}
\DontPrintSemicolon
\SetAlgoLined

\SetKwInOut{Input}{Input}
\SetKwInOut{Output}{Output}

\Input{$N, K, \textcolor{black}{N_{anc}}, \mathcal{T}_{\text{sl}}, \text{max\_iter}, S$}
\Output{Optimized pilot set $\mathcal{P}_{\text{opt}}$}

\textcolor{black}{$\mathcal{P} \gets \text{Scattered anchors of size } N_{anc}$}\;
\textcolor{black}{$\mathcal{U} \gets \{0,\dots,N-1\} \setminus \mathcal{P}$}\;

\For{\textcolor{black}{$k = N_{anc} + 1$} \textbf{to} $K$}{
    $n^* \gets \arg\min\limits_{n \in \mathcal{U}} \max\limits_{\tau \in \mathcal{T}_{\text{sl}}} |\Psi\textcolor{black}{_{pp}}(\tau; \mathcal{P} \cup \{n\})|$\;
    $\mathcal{P} \gets \mathcal{P} \cup \{n^*\}$, $\mathcal{U} \gets \mathcal{U} \setminus \{n^*\}$\;
}

$\text{iter} \gets 0$, $\text{improved} \gets \text{True}$\;

\While{\emph{improved} \textbf{and} $\text{iter} < \text{max\_iter}$}{
    $\text{improved} \gets \text{False}$\;
    \For{\textbf{each} $p \in \mathcal{P}$ \textbf{in random order}}{
        $\mathcal{U}_S \gets$ random subset of $\mathcal{U}$ with size $S$\;
        \For{\textbf{each} $u \in \mathcal{U}_S$}{
            $\mathcal{P}' \gets (\mathcal{P} \setminus \{p\}) \cup \{u\}$\;
            \If{$\mathrm{PSL}(\mathcal{P}') < \mathrm{PSL}(\mathcal{P})$}{
                $\mathcal{P} \gets \mathcal{P}'$, $\mathcal{U} \gets (\mathcal{U} \setminus \{u\}) \cup \{p\}$\;
                $\text{improved} \gets \text{True}$\;
                \textbf{break}\;
            }
        }
    }
    $\text{iter} \gets \text{iter} + 1$\;
}
\Return $\mathcal{P}$\;

\end{algorithm}

\begin{figure*}[htbp]
\centering
\includegraphics[width=1\textwidth]{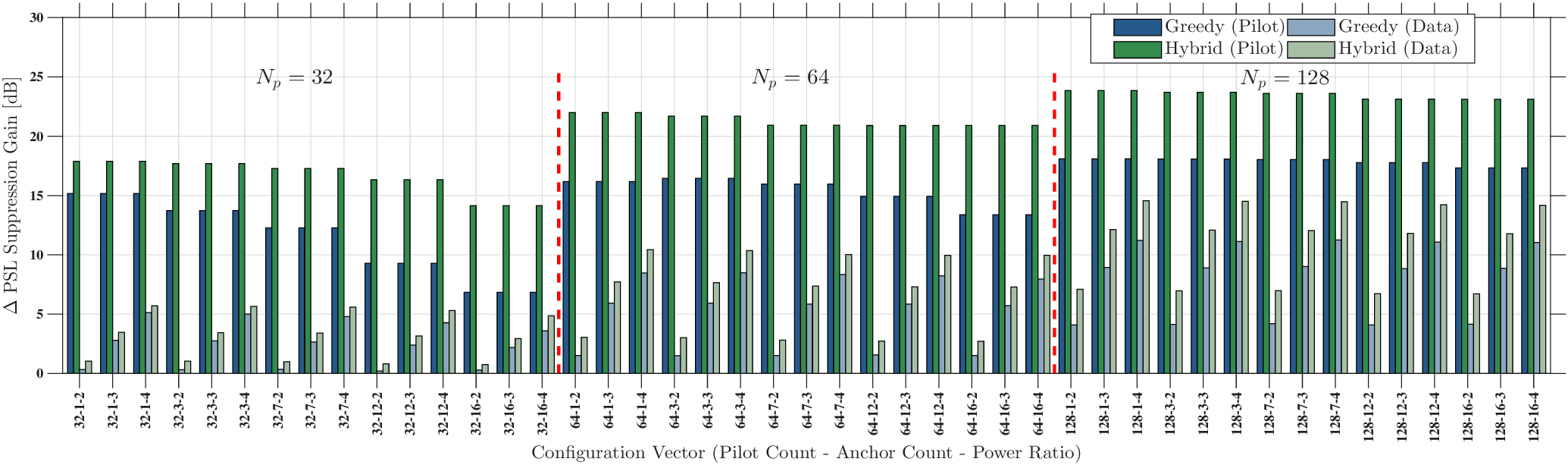} 
\caption{\textcolor{black}{Theoretical vs. Operational PSL Suppression Gain ($\Delta$PSL) evaluation across varying pilot densities and anchor constraints.}}
\label{fig:dual_bound_ablation}
\end{figure*}
\section{Numerical Validation}
\label{sec:validation}

We validate the proposed framework using an \ac{OFDM} system with $N=512$ subcarriers. The performance is evaluated using 16-QAM data modulation and full waveform simulation. \textcolor{black}{To explicitly evaluate the sensing-communication trade-off, we enforce structural communication constraints by locking $N_{anc}$ widely distributed anchor pilots (e.g., $N_{anc} \in \{1, 5, 12, 16\}$) to guarantee channel estimation coverage, the pilot-to-data power ratio is set to 4.}
\textcolor{black}{To quantify the magnitude of the optimization, Fig.~\ref{fig:dual_bound_ablation} evaluates the \ac{PSL} Suppression Gain ($\Delta\mathrm{PSL}$) relative to the standard periodic baseline, mathematically defined as:
\begin{equation}
    \Delta\mathrm{PSL} (\mathrm{dB}) = \mathrm{PSL}_{\mathrm{periodic}} (\mathrm{dB}) - \mathrm{PSL}_{\mathrm{evaluated}} (\mathrm{dB}).
\end{equation}
A positive $\Delta\mathrm{PSL}$ indicates a reduction in the grating lobe floor. We analyze performance under two distinct bounds. We analyze performance under two distinct bounds. In the theoretical pilot only condition, unallocated subcarriers contain zero energy. Under strict anchor constraints, the Greedy algorithm suffers from local minima. Conversely, the Hybrid \ac{SCCD} spatial redistribution breaks the induced periodicities, yielding theoretical suppression gains exceeding 5 dB for sparse densities ($N_p=32$) where $N_p$ (equivalent to $K$) denotes the pilot count.}

\textcolor{black}{In the operational data included condition, unallocated subcarriers transmit independent, unit-variance \ac{QAM} data ($P_{ratio}=4$). This introduces a stochastic \ac{DPI} noise floor, which physically bounds the absolute \ac{PSL}. However, as shown in Fig.~\ref{fig:dual_bound_ablation}, the Hybrid pattern structurally suppresses the deterministic pilot grating lobes ($\Psi_{pp}$) deep enough that the remaining interference is strictly dominated by the communication noise floor, maintaining performance advantage over the Greedy baseline across all evaluated configurations.}

\textcolor{black}{Standard periodic (Uniform) patterns exhibit a terminal \ac{RMSE} saturation floor, as shown in Fig.~\ref{fig:rmse}. This floor persists regardless of \ac{SNR} because deterministic pilot grating lobes induce ghost targets that scale linearly with signal power. While the Greedy approach attempts suppression, it is limited by constraint saturation when rigidly locked to $N_{anc}=16$ anchors. The proposed Hybrid \ac{SCCD} algorithm breaks these periodicities, maintaining a noise-limited, monotonically decreasing \ac{RMSE} curve across the evaluated \ac{SNR}.}

\textcolor{black}{Unconstrained pilot optimization leads to pilot clustering, which violates the Nyquist sampling theorem in the frequency domain and results in a poor communication link. As demonstrated in Fig.~\ref{fig:ber}, by enforcing a rigid distribution of $N_{anc}=16$ anchors, the wideband channel remains sampled at intervals shorter than the coherence bandwidth. This structural constraint perfectly preserves the channel estimation quality, yielding a \ac{BER} curve that strictly overlaps with the standard periodic baseline, effectively decoupling sensing enhancement from communication reliability.}


\section{Conclusion}
\label{sec:conclusion}
\textcolor{black}{This paper presented a hybrid optimization framework for \ac{ISAC} pilot design. We demonstrated that while standard constructive algorithms provide an initial improvement over periodic patterns, they are severely limited by local minima when subjected to practical communication constraints. By augmenting a greedy initialization with stochastic cyclic coordinate descent (\ac{SCCD}) refinement, the proposed hybrid algorithm successfully shatters cyclic periodicities. System-level evaluations confirm that this structural redistribution resolves the fundamental sensing-communication trade-off. It suppresses deterministic grating lobes beneath the impassable data-to-pilot interference noise floor, effectively eliminating the terminal range \ac{RMSE} saturation inherent to periodic grids, without degrading the primary communication \ac{BER}. Future work will address the design of efficient signaling schemes to convey optimized pilot indices and the investigation of joint receiver-side processing techniques.}

\IEEEpeerreviewmaketitle
\bibliographystyle{IEEEtran}
\bibliography{IEEEfull.bib}

\vfill
\end{document}

%% file: acronyms.tex
\begin{acronym}
  \acro{SCS}{Spectally-Constrained Sequences}
  \acro{CDS}{Cyclic Difference Sets}
  \acro{SCCD}{stochastic cyclic coordinate descent}
  \acro{PSL}{peak to sidelobe level}
  \acro{3GPP}{3rd Generation Partnership Project}
  \acro{4G}{fourth generation}
  \acro{5G}{fifth generation}
  \acro{5G-NR}{5G New Radio}
  \acro{6G}{sixth generation}
  \acro{AWGN}{additive white Gaussian noise}
  \acro{BW}{bandwidth}
  \acro{AF}{ambiguity function}
  \acro{PHY}{physical-layer}
  \acro{LUT}{Look-up Table}
  \acro{RF}{radio frequency}
  \acro{CFO}{carrier frequency offset}
  \acro{CP}{cyclic prefix}
  \acro{OFDM}{orthogonal frequency-division multiplexing}
  \acro{CRLB}{Cramér--Rao lower bound}
  \acro{DAC}{digital-to-analog converter}
  \acro{DC}{direct current}
  \acro{CPI}{coherent processing interval}
  \acro{DFT}{discrete Fourier transform}
  \acro{DL}{downlink}
  \acro{EVD}{eigenvalue decomposition}
  \acro{FFT}{fast Fourier transform}
  \acro{FMCW}{frequency-modulated continuous wave}
  \acro{FoV}{field of view}
  \acro{FR}{frequency range}
  \acro{IDFT}{inverse discrete Fourier transform}
  \acro{IFFT}{inverse fast Fourier transform}
  \acro{IoT}{Internet of Things}
  \acro{ISAC}{Integrated sensing and communication}
  \acro{ISI}{inter-symbol interference}
  \acro{CRB}{ cramér–rao bound}
  \acro{JD}{joint design}
  \acro{JSAC}{joint sensing and communication}
  \acro{LEO}{low Earth orbit}
  \acro{LLR}{log-likelihood ratio}
  \acro{LOS}{line of sight}
  \acro{LTE}{Long Term Evolution}
  \acro{NMSE}{normalized mean squared error}
  \acro{OFDM}{orthogonal frequency-division                    multiplexing}
  \acro{QAM}{quadrature amplitude modulation}
  \acro{QPSK}{Quadrature Phase Shift Keying }
  \acro{MF}{matched filter}
  \acro{RCS}{radar cross section}
  \acro{CDF}{cumulative distribution function}
  \acro{SNR}{signal to noise ratio}
  \acro{RMSE}{root mean squared error}
  \acro{BER}{bit error rate}
  \acro{DPI}{data-to-pilot interference}

\end{acronym}